\documentclass[reqno]{amsart}
\usepackage{amsmath, euscript, amsthm}

\font\msbm=msbm10
\font\msbmsmall=msbm10 scaled 800
\font\msbmtiny=msbm10 scaled 700
\newfam\msbmfam
\textfont\msbmfam=\msbm
\scriptfont\msbmfam=\msbmsmall
\scriptscriptfont\msbmfam=\msbmtiny

\def\diag{\mbox{\rm diag}}

\def\varkappa{\mbox{\msbm \char'173}}      

\def\Bbb{\mathbb}
\def\cal{\mathcal}

\theoremstyle{plain}

\newtheorem{RHP}{Riemann-Hilbert problem}

\theoremstyle{definition}

\theoremstyle{remark}

\numberwithin{equation}{section}

\numberwithin{figure}{section}

\begin{document}
\title[The RH problem for the bi-orthogonal polynomials]
{The Riemann-Hilbert Problem for the Bi-Orthogonal Polynomials}
\author{Andrei A. Kapaev}
\address{St Petersburg Department of Steklov Mathematical Institute
of Russian Academy of Sciences, Fontanka 27, St Petersburg, 191011, Russia}
\curraddr{Department of Applied Mathematics and Theoretical Physics,
University of Cambridge, Silver st., Cambridge, CB3 9EW, England}
\email{kapaev@pdmi.ras.ru, A.Kapaev@damtp.cam.ac.uk}

\thanks{This work was partially supported by the EPSRC and by RFBR (grant
No~02-01-00268)}
\thanks{The author thanks J.~Harnad and M.~Bertola for stimulating
discussions and valuable comments. The author is also grateful to
A.~S.~Fokas for support and to the staff of DAMTP, University of
Cambridge, for the hospitality during his visit}

\begin{abstract}
For the bi-orthogonal polynomials with the third degree polynomial 
potential functions, the $3\times3$ matrix Riemann-Hilbert problem is 
explicitly constructed. The developed approach admits an extension to the 
bi-orthogonal polynomials with arbitrary polynomial potentials.
\end{abstract}
\maketitle

\section{Introduction}

The classical asymptotic theory of the orthogonal polynomials 
\cite{PR, szego} extended to the polynomials orthogonal w.r.t.\ the weight
$e^{-NV(z)}$ \cite{F, N, M, Sh} has gained an essential progress after
introduction of the Riemann-Hilbert (RH) problem approach \cite{FIK} and 
the steepest descent method \cite{DZ}. A particular implementation of 
these methods to the orthogonal polynomials on the real line can by found 
in \cite{BI, DKMcLVZ}; a similar study of the orthogonal polynomials on the 
circle is given in \cite {BDJ}.

Further extensions of the notion of the orthogonal polynomials motivated 
by a number of applications to a random matrix theory, integrable systems,
approximation theory and combinatorics include the generalized orthogonal 
polynomials and the bi-orthogonal polynomials. In the former case, the
sequence of the polynomials is orthogonal w.r.t.\ a sequence of measures 
\cite{AvM}, 
$$
\int_{\Bbb R}P_n(\lambda)\,d\rho_m(\lambda)=\delta_{nm},
$$
while in the latter case, two sequences of the polynomials are 
orthogonal to each other w.r.t.\ a two-dimensional measure, see 
\cite{EMcL, BEH, Z-J},
$$
\int_{\Bbb R}\int_{\Bbb R}
P_n(\lambda)Q_m(\xi)\,d\mu(\lambda,\xi)=\delta_{nm}.
$$
In the simplest case, the two-dimensional measure has the form of the 
product $d\mu(\lambda,\xi)=
e^{-V(\lambda)-W(\xi)+\lambda\xi}\,d\lambda\,d\xi$ 
where the polynomials $V(\lambda)$ and $W(\xi)$ are called {\em potentials}. 
Integration over $\xi$ in the latter two-fold integral yields a sequence of 
measures, $d\rho_m(\lambda)=\hat\phi_m(\lambda)\,d\lambda
=\int_{\Bbb R}Q_m(\xi)\,d\mu(\lambda,\xi)$. The functions
$\hat\phi_m(\lambda)$ here are called the {\em dual functions}, 
see \cite{BEH}. 

The generalized and bi-orthogonal polynomials are associated to a completely 
integrable system coming from the $t$-deformations and the Virasoro 
constraints and describing typically the reductions of the 2-Toda lattice, 
see \cite{AvM, vM, EMcL, BEH, Z-J}. However, in spite of the algebraic 
properties of the generalized and bi-orthogonal polynomials are well known, 
the knowledge of the asymptotic properties of such polynomials is very 
limited. Basically, this is due to the absence of an adequate formulation 
of the relevant RH problem. Indeed, in spite of the $2\times2$ matrix RH 
problem formulation for the conventional orthogonal polynomials \cite{FIK} 
admits a direct $2\times2$ extension for the generalized orthogonal 
polynomials \cite{AvM}, the relevant version for the bi-orthogonal 
polynomials \cite{EMcL} shows its non-local nature. 

Fortunately, the study \cite{BEH} reveals the isomonodromy structure
associated to the bi-orthogonal polynomials and hence the principal 
possibility to formulate an $n\times n$ matrix RH problem which would enjoy 
the properties similar to those for the $2\times2$ matrix RH problem for 
the conventional orthogonal polynomials. In what follows, we construct the 
$3\times3$ matrix RH problem for the bi-orthogonal polynomials with the 
third degree polynomial potentials, as well as the $3\times3$ RH problem 
for the relevant dual functions. In spite of its less physical 
importance, this case provides us with the opportunity to develop the 
technique in the simplest non-trivial case. Indeed, the bi-orthogonal 
polynomials for the second degree potentials are reduced to the classical 
Hermite polynomials \cite{EMcL}. In some more involved case 
$\deg V(\lambda)>2$, $\deg W(\xi)=2$, the bi-orthogonal polynomials can be 
expressed in terms of the semi-classical orthogonal polynomials related to 
the $2\times2$ matrix RH problem studied in \cite{FIK, BI, BDJ, DKMcLVZ} 
and other papers. 

We stress that the method presented below can be extended to arbitrary 
polynomial potentials. We also point out a particular importance of the 
paper \cite{B} which is useful for construction and justification of the 
RH problem for the dual functions for the class of the weights with the
rational log-derivatives. After the original version of the present
paper was posted to the internet, the Montreal group presented their
methodology for construction similar RH problem \cite{BEH2}. Their
idea for evaluation of the RH problem for the dual functions based on
the study of the path integrals coincides with the presented below
but, in contrast to our cubic case, they consider the arbitrary
polynomial potentials. As to the RH problem for the original (wave)
function, the authors of \cite{BEH2} rely on the so-called ``duality
pairing'', while our approach is based on the explicit integral
representation for the wave function.

The paper is organized as follows. In Section~\ref{diff_eqs}, we
recall the matrix differential and difference equations satisfied by 
the bi-orthogonal polynomials and their dual functions. In
Section~\ref{dual_RH}, we construct particular matrix solutions for the
differential-difference systems for the dual functions which gives rise 
to a RH problems for the dual functions
(\ref{hat_Psi_n_as_RH})--(\ref{hat_Psi_branch_jump}) and
(\ref{hat_Phi_m_as_RH})--(\ref{hat_Phi_branch_jump}). 
In Section~\ref{wave_RH}, we construct particular solutions of the 
differential-difference systems for the bi-orthogonal polynomials and 
the relevant RH problems (\ref{Psi_n_as_RH})--(\ref{Psi_branch_jump}) 
and (\ref{Phi_m_as_RH})--(\ref{Phi_branch_jump}).

\section{Equations for the bi-orthogonal polynomials}\label{diff_eqs}

Below, we consider the monic polynomials $p_n(\lambda)$, $q_m(\xi)$ 
satisfying the orthogonality condition 
\begin{equation}\label{ortho_condition}
\int_{\gamma_1}d\lambda\int_{\gamma_2}d\xi\,\,p_n(\lambda)q_m(\xi)
e^{-V(\lambda)-W(\xi)+t\lambda\xi}=h_n^2\delta_{n,m},
\end{equation}
where
\begin{equation}\label{VW_3}
V(\lambda)=\tfrac{1}{3}\lambda^3+x\lambda,\quad
W(\xi)=\tfrac{1}{3}\xi^3+y\xi,
\end{equation}
$x,y,t\in{\Bbb C}$, $t\neq0$; the contours $\gamma_i$, $i=1,2$, are
the complex linear combinations of the elementary contours, 
\begin{equation}\label{gamma_def}
\gamma_i=\sum_{j=0}^2g_j^{(i)}\Gamma_j,\quad
g_j^{(i)}\in{\Bbb C},\quad
i=1,2,
\end{equation}
where each $\Gamma_j$ is the sum of two rays,
\begin{equation}\label{Gamma_def}
\Gamma_j=(e^{i\frac{2\pi}{3}(j-2)}\infty,0]\cup
[0,e^{i\frac{2\pi}{3}(j-1)}\infty),\quad 
j=0,1,2.
\end{equation}
Because $\Gamma_0+\Gamma_1+\Gamma_2=0$, one of the parameters
$g_j^{(1)}$ (resp., $g_j^{(2)}$) can be put to zero; one of the nontrivial 
parameters $g_j^{(i)}$ can be normalized to the unit. Thus the set of 
contours $\gamma_j$ and therefore the set of the monic bi-orthogonal 
polynomials is parameterized by three constant complex
parameters\footnote{More general parameterization of the set of
contours is introduced in \cite{BEH2}, however, in our cubic case,
both kinds of parameterizations are equivalent}.

Introduce the wave functions 
\begin{equation}\label{psi_phi_def}
\psi_n(\lambda)=
\frac{1}{h_n}p_n(\lambda)e^{-V(\lambda)},\quad
\phi_m(\xi)=
\frac{1}{h_m}q_m(\xi)e^{-W(\xi)},
\end{equation}
and define the dual functions $\hat\psi_n(\xi)$,
$\hat\phi_m(\lambda)$ which are the Fourier-Laplace images of the
wave functions \cite{BEH},
\begin{equation}\label{dual_psi_phi}
\hat\psi_n(\xi)=
\int_{\gamma_1}\psi_n(\lambda)e^{t\lambda\xi}\,d\lambda,\quad
\hat\phi_m(\lambda)=
\int_{\gamma_2}\phi_m(\xi)e^{t\lambda\xi}\,d\xi.
\end{equation}

The orthogonality condition (\ref{ortho_condition}) now reads
\begin{equation}\label{ortho_norm_condition}
\int_{\gamma_1}d\lambda\int_{\gamma_2}d\xi\,
\psi_n(\lambda)\phi_m(\xi)e^{t\lambda\xi}=
\int_{\gamma_1}
\psi_n(\lambda)\hat\phi_m(\lambda)\,d\lambda=
\int_{\gamma_2}
\hat\psi_n(\xi)\phi_m(\xi)\,d\xi=\delta_{nm}\,.
\end{equation}
It implies certain relations between the introduced functions
(\ref{psi_phi_def}), (\ref{dual_psi_phi}). We refer \cite{BEH} for the 
general case and present here the final result for our particular 
situation:
\begin{equation}\label{eqs_psi}
\begin{split}
&\lambda\psi_n(\lambda)=\sum_{m=n-2}^{n+1}a_{n,m}\psi_m(\lambda),\quad
\partial_{\lambda}\psi_n(\lambda)=
-t\sum_{m=n-1}^{n+2}b_{n,m}\psi_m(\lambda),
\\
&\partial_x\psi_n(\lambda)=\sum_{m=n}^{n+1}u_{n,m}\psi_m(\lambda),\quad
\partial_y\psi_n(\lambda)=-\sum_{m=n-1}^nv_{n,m}\psi_m(\lambda),
\\
&\partial_t\psi_n(\lambda)=w_n\psi_n(\lambda)
-\sum_{m=n-3}^{n-1}A_{n,m}\psi_m(\lambda),
\end{split}
\end{equation}
\begin{equation}\label{eqs_phi}
\begin{split}
&\xi\phi_m(\xi)=\sum_{n=m-2}^{m+1}b_{n,m}\phi_n(\xi),\quad
\partial_{\xi}\phi_m(\xi)=-t\sum_{n=m-1}^{m+2}a_{n,m}\phi_n(\xi),
\\
&\partial_y\phi_m(\xi)=\sum_{n=m}^{m+1}v_{n,m}\phi_n(\xi),\quad
\partial_x\phi_m(\xi)=-\sum_{n=m-1}^mu_{n,m}\phi_n(\xi),
\\
&\partial_t\phi_m(\xi)=w_m\phi_m(\xi)
-\sum_{n=m-3}^{m-1}A_{n,m}\phi_n(\xi),
\end{split}
\end{equation}
where
$$
a_{n,n+1}=b_{n+1,n}=-u_{n,n+1}=-v_{n+1,n}=\frac{h_{n+1}}{h_n}\,,\quad
a_{n+2,n}=b_{n,n+2}=\frac{h_{n+2}}{t\,h_n};
$$
$$
u_{n,n}=\partial_x\ln h_n,\quad
v_{n,n}=\partial_y\ln h_n,\quad
A_{n,n}=-2w_n=2\partial_t\ln h_n;
$$
$$
A_{n,m}=
\begin{cases}
\sum_{k=n-2}^{m+1}a_{n,k}b_{k,m},\quad
0\leq n-m\leq3,\\
\sum_{k=m-2}^{n+1}a_{n,k}b_{k,m},\ 
-3\leq n-m\leq0.
\end{cases}
$$
Using the definition (\ref{dual_psi_phi}) and the equations above, it is
straightforward that the dual functions satisfy 
\begin{equation}\label{dual_eqs_psi}
\begin{split}
&\partial_{\xi}\hat\psi_n(\xi)=
\sum_{m=n-2}^{n+1}ta_{n,m}\hat\psi_m(\xi),\quad
\xi\hat\psi_n(\xi)=
\sum_{m=n-1}^{n+2}b_{n,m}\hat\psi_m(\xi),
\\
&\partial_x\hat\psi_n(\xi)=
\sum_{m=n}^{n+1}u_{n,m}\hat\psi_m(\xi),\quad
\partial_y\hat\psi_n(\xi)=
-\sum_{m=n-1}^nv_{n,m}\hat\psi_m(\xi),
\\
&\partial_t\hat\psi_n(\xi)=
-w_n\hat\psi_n(\xi)
+\sum_{m=n+1}^{n+3}A_{n,m}\hat\psi_m(\xi),
\end{split}
\end{equation}
\begin{equation}\label{dual_eqs_phi}
\begin{split}
&\partial_{\lambda}\hat\phi_m(\lambda)=
\sum_{n=m-2}^{m+1}tb_{n,m}\hat\phi_n(\lambda),\quad
\lambda\hat\phi_m(\lambda)=
\sum_{n=m-1}^{m+2}a_{n,m}\hat\phi_n(\lambda),
\\
&\partial_y\hat\phi_m(\lambda)=
\sum_{n=m}^{m+1}v_{n,m}\hat\phi_n(\lambda),\quad
\partial_x\hat\phi_m(\lambda)=
-\sum_{n=m-1}^mu_{n,m}\hat\phi_n(\lambda),
\\
&\partial_t\hat\phi_m(\lambda)=
-w_m\hat\phi_m(\lambda)
+\sum_{n=m+1}^{m+3}A_{n,m}\hat\phi_n(\lambda).
\end{split}
\end{equation}

The above expressions can be written in the matrix form \cite{BEH}. Indeed,
the $3$-vector $\Psi_n(\lambda)=\bigl(\psi_n(\lambda),\psi_{n-1}(\lambda),
\psi_{n-2}(\lambda)\bigr)^T$ satisfies the system of difference and
differential equations,
\begin{equation}\label{matrix_eqs_Psi}
\begin{split}
&\Psi_{n+1}(\lambda)=R_n(\lambda)\Psi_n(\lambda),\quad
\frac{\partial\Psi_n}{\partial\lambda}(\lambda)=
A_n(\lambda)\Psi_n(\lambda),
\\
&\frac{\partial\Psi_n}{\partial x}=U_n(\lambda)\Psi_n,\quad
\frac{\partial\Psi_n}{\partial y}=V_n(\lambda)\Psi_n,\quad
\frac{\partial\Psi_n}{\partial t}=W_n(\lambda)\Psi_n,
\end{split}
\end{equation}
where
$$
R_n(\lambda)=
\begin{pmatrix}
\frac{\lambda-a_{n,n}}{a_{n,n+1}}&
-\frac{a_{n,n-1}}{a_{n,n+1}}&
-\frac{a_{n,n-2}}{a_{n,n+1}}\\
1&0&0\\
0&1&0
\end{pmatrix},
$$
$$
A_n(\lambda)=-D_{n,n+2}R_{n+1}(\lambda)R_n(\lambda)
-D_{n,n+1}R_{n}(\lambda)
-D_{n,n}
-D_{n,n-1}R_{n-1}^{-1}(\lambda),
$$
$$
D_{n,m}=t\,\diag\bigl(
b_{n,m},b_{n-1,m-1},b_{n-2,m-2}\bigr),
$$
$$
U_n(\lambda)=U_{n,n+1}R_n(\lambda)+U_{n,n},\quad
U_{n,m}=\diag\bigl(
u_{n,m},u_{n-1,m-1},u_{n-2,m-2}\bigr),
$$
$$
V_n(\lambda)=-V_{n,n}-V_{n,n-1}R_{n-1}^{-1}(\lambda),\quad
V_{n,m}=\diag\bigl(
v_{n,m},v_{n-1,m-1},v_{n-2,m-2}\bigr),
$$
\begin{multline*}
W_n(\lambda)=W_{n,n}-{\cal A}_{n,n-1}R_{n-1}^{-1}(\lambda)
-{\cal A}_{n,n-2}R_{n-2}^{-1}(\lambda)R_{n-1}^{-1}(\lambda)-
\\
-{\cal A}_{n,n-3}R_{n-3}^{-1}(\lambda)R_{n-2}^{-1}(\lambda)
R_{n-1}^{-1}(\lambda),
\end{multline*}
$$
W_{n,n}=\diag\bigl(w_n,w_{n-1},w_{n-2}\bigr),\quad
{\cal A}_{n,m}=\diag\bigl(A_{n,m},A_{n-1,m-1},A_{n-2,m-2}\bigr).
$$

Similarly, the $3$-vector 
$\Phi_m(\xi)=\bigl(\phi_m(\xi),\phi_{m-1}(\xi),\phi_{m-2}(\xi)\bigr)^T$ 
satisfies the system
\begin{equation}\label{matrix_eqs_Phi}
\begin{split}
&\Phi_{m+1}(\xi)=Q_m(\xi)\Phi_m(\xi),\quad
\frac{\partial\Phi_m}{\partial\xi}(\xi)=B_m(\xi)\Phi_m(\xi),
\\
&\frac{\partial\Phi_m}{\partial x}={\cal U}_m\Phi_m,\quad
\frac{\partial\Phi_m}{\partial y}={\cal V}_m\Phi_m,\quad
\frac{\partial\Phi_m}{\partial t}={\cal W}_m\Phi_m,
\end{split}
\end{equation}
where
$$
Q_m(\xi)=\begin{pmatrix}
\frac{\xi-b_{m,m}}{b_{m+1,m}}&
-\frac{b_{m-1,m}}{b_{m+1,m}}&
-\frac{b_{m-2,m}}{b_{m+1,m}}
\\
1&0&0\\
0&1&0
\end{pmatrix},
$$
$$
B_m(\xi)=-{\cal D}_{m+2,m}Q_{m+1}(\xi)Q_{m}(\xi)
-{\cal D}_{m+1,m}Q_{m}(\xi)
-{\cal D}_{m,m}
-{\cal D}_{m-1,m}Q_{m-1}^{-1}(\xi),
$$
$$
{\cal D}_{m,n}=t\,\diag\bigl(a_{m,n},a_{m-1,n-1},a_{m-2,n-2}\bigr),
$$
$$
{\cal U}_m(\xi)=-U_{m,m}-U_{m-1,m}Q_{m-1}^{-1}(\xi),\quad
{\cal V}_m(\xi)=V_{m+1,m}Q_m(\xi)+V_{m,m},
$$
\begin{multline*}
{\cal W}_m(\xi)=W_{m,m}-{\cal A}_{m-1,m}Q_{m-1}^{-1}(\xi)-
{\cal A}_{m-2,m}Q_{m-2}^{-1}(\xi)Q_{m-1}^{-1}(\xi)-
\\
-{\cal A}_{m-3,m}Q_{m-3}^{-1}(\xi)Q_{m-2}^{-1}(\xi)Q_{m-1}^{-1}(\xi).
\end{multline*}

The matrix equations for the $3$-vector of dual functions,
$$
\hat\Psi_n(\xi)=
\bigl(\hat\psi_n(\xi),\hat\psi_{n-1}(\xi),\hat\psi_{n-2}(\xi)\bigr)^T,$$
are as follows,
\begin{equation}\label{matrix_eqs_dual_Psi}
\begin{split}
&\hat\Psi_{n+1}(\xi)=\hat R_n(\xi)\hat\Psi_n(\xi),\quad
\frac{\partial\hat\Psi_n}{\partial\xi}(\xi)=
\hat A_n(\xi)\hat\Psi_n(\xi),
\\
&\frac{\partial\hat\Psi_n}{\partial x}(\xi)=
\hat U_n(\xi)\hat\Psi_n(\xi),\quad
\frac{\partial\hat\Psi_n}{\partial y}(\xi)=
\hat V_n(\xi)\hat\Psi_n(\xi),\quad
\frac{\partial\hat\Psi_n}{\partial t}(\xi)=
\hat W_n(\xi)\hat\Psi_n(\xi),
\end{split}
\end{equation}
where
$$
\hat R_n(\xi)=
\begin{pmatrix}
-\frac{b_{n-1,n}}{b_{n-1,n+1}}&
\frac{\xi-b_{n-1,n-1}}{b_{n-1,n+1}}&
-\frac{b_{n-1,n-2}}{b_{n-1,n+1}}\\
1&0&0\\
0&1&0
\end{pmatrix},
$$
$$
\hat A_n(\xi)={\cal D}_{n,n+1}\hat R_n(\xi)+{\cal D}_{n,n}
+{\cal D}_{n,n-1}\hat R_{n-1}^{-1}(\xi)
+{\cal D}_{n,n-2}\hat R_{n-2}^{-1}(\xi)\hat R_{n-1}^{-1}(\xi),
$$
$$
\hat U_n(\xi)=U_{n,n+1}\hat R_n(\xi)+U_{n,n},\quad
\hat V_n(\xi)=-V_{n,n}-V_{n,n-1}\hat R_{n-1}^{-1}(\xi),
$$
\begin{multline*}
\hat W_n(\xi)=-W_{n,n}
+{\cal A}_{n,n+3}\hat R_{n+2}(\xi)\hat R_{n+1}(\xi)\hat R_n(\xi)+
\\
+{\cal A}_{n,n+2}\hat R_{n+1}(\xi)\hat R_n(\xi)
+{\cal A}_{n,n+1}\hat R_n(\xi).
\end{multline*}

In the very same way, the $3$-vector 
$$
\hat\Phi_m(\lambda)=
\bigl(\hat\phi_m(\lambda),\hat\phi_{m-1}(\lambda),
\hat\phi_{m-2}(\lambda)\bigr)^T
$$
satisfies the system of equations
\begin{equation}\label{matrix_eqs_dual_Phi}
\begin{split}
&\hat\Phi_{m+1}(\lambda)=\hat Q_m(\lambda)\hat\Phi_m(\lambda),\quad
\frac{\partial\hat\Phi_m}{\partial\lambda}(\lambda)=
\hat B_m(\lambda)\hat\Phi_m(\lambda),
\\
&\frac{\partial\hat\Phi_m}{\partial x}(\lambda)=
\hat{\cal U}_m(\lambda)\hat\Phi_m(\lambda),\quad
\frac{\partial\hat\Phi_m}{\partial y}(\lambda)=
\hat{\cal V}_m(\lambda)\hat\Phi_m(\lambda),\quad
\frac{\partial\hat\Phi_m}{\partial t}(\lambda)=
\hat{\cal W}_m(\lambda)\hat\Phi_m(\lambda),
\end{split}
\end{equation}
where
$$
\hat Q_m(\lambda)=
\begin{pmatrix}
-\frac{a_{m,m-1}}{a_{m+1,m-1}}&
\frac{\lambda-a_{m-1,m-1}}{a_{m+1,m-1}}&
-\frac{a_{m-2,m-1}}{a_{m+1,m-1}}\\
1&0&0\\
0&1&0
\end{pmatrix},
$$
$$
\hat B_m(\lambda)=
D_{m+1,m}\hat Q_m(\lambda)+D_{m,m}
+D_{m-1,m}\hat Q_{m-1}^{-1}(\lambda)
+D_{m-2,m}\hat Q_{m-2}^{-1}(\lambda)\hat Q_{m-1}^{-1}(\lambda),
$$
$$
\hat{\cal U}_m(\lambda)=
-U_{m,m}-U_{m-1,m}\hat Q_{m-1}^{-1}(\lambda),\quad
\hat{\cal V}_m(\lambda)=V_{m+1,m}\hat Q_m(\lambda)+V_{m,m},
$$
\begin{multline*}
\hat{\cal W}_m(\lambda)=-W_{m,m}
+{\cal A}_{m+3,m}
\hat Q_{m+2}(\lambda)\hat Q_{m+1}(\lambda)\hat Q_m(\lambda)+
\\
+{\cal A}_{m+2,m}\hat Q_{m+1}(\lambda)\hat Q_m(\lambda)
+{\cal A}_{m+1,m}\hat Q_m(\lambda).
\end{multline*}

The compatibility conditions of the above equations yield a reduction 
of the 2-Toda lattice, see \cite{AvM}. On the other hand, this nonlinear 
system describes the isomonodromy deformations of the $3\times3$ matrix 
differential equations in $\lambda$ and $\xi$. The matrix solutions of 
the systems (\ref{matrix_eqs_Psi}), (\ref{matrix_eqs_Phi}), 
(\ref{matrix_eqs_dual_Psi}) and (\ref{matrix_eqs_dual_Phi}) give rise to 
a matrix RH problems whose data can be found via the asymptotic analysis 
of the $\lambda$- and $\xi$-equations with the polynomial matrices 
$A_n(\lambda)$, $B_m(\xi)$ and $\hat A_n(\lambda)$, $\hat B_m(\xi)$ using 
the WKB technique, see \cite{IN}. It is crucial however, that it is 
possible to avoid the WKB analysis expanding the basic idea of \cite{FIK}. 
In what follows, we construct certain particular solutions of the above 
systems in terms of the (unknown) bi-orthogonal polynomials using merely 
the fact of the existence of the linear systems (\ref{matrix_eqs_Psi}), 
(\ref{matrix_eqs_Phi}), (\ref{matrix_eqs_dual_Psi}), 
(\ref{matrix_eqs_dual_Phi}) rather than the systems themselves.

\section{The particular solutions of the matrix equations and 
the Riemann-Hilbert problem for the dual functions}\label{dual_RH}

Because of the mentioned above independence of the relevant monodromy
data on the deformation parameter $t$, without loss of generality, we
restrict ourselves to $t>0$. This assumption allows us to simplify our
calculations while the final result will be valid for arbitrary
$t\in{\Bbb C}\backslash\{0\}$.

Let us introduce the auxiliary functions
\begin{equation}\label{aux_def}
\hat\psi_n^{(j)}(\xi)=
\int_{\Gamma_j}\psi_n(\lambda)e^{t\lambda\xi}\,d\lambda,\quad
\hat\phi_m^{(j)}(\lambda)=
\int_{\Gamma_j}\phi_m(\xi)e^{t\lambda\xi}\,d\xi,
\end{equation}
where the contours $\Gamma_j$, $j=0,1,2$, are defined in
(\ref{Gamma_def}). Due to (\ref{dual_psi_phi}) and (\ref{gamma_def}),
the dual and auxiliary functions are related to each other via
\begin{equation}\label{dual_via_aux}
\hat\psi_n(\xi)=\sum_{j=0}^2g_j^{(1)}\hat\psi_n^{(j)}(\xi),\quad
\hat\phi_m(\lambda)=\sum_{j=0}^2g_j^{(2)}\hat\phi_m^{(j)}(\lambda),
\end{equation}
and satisfy the same equations (\ref{dual_eqs_psi}), (\ref{dual_eqs_phi}).
Taking into account the orthogonality condition
(\ref{ortho_norm_condition}), we find that the functions
\begin{equation}\label{F_n_def}
F_n(\xi)=\frac{e^{W(\xi)}}{2\pi i}\int_{\gamma_2}
\frac{\hat\psi_n(\zeta)e^{-W(\zeta)}}{\zeta-\xi}\,d\zeta,\quad
G_m(\lambda)=\frac{e^{V(\lambda)}}{2\pi i}\int_{\gamma_1}
\frac{\hat\phi_m(\zeta)e^{-V(\zeta)}}{\zeta-\lambda}\,d\zeta,
\end{equation}
for $n,m\geq2$ also satisfy the same equations.

Observing that $\hat\psi_n^{(j)}$, $j=1,2$, (respectively,
$\hat\phi_m^{(j)}$, $j=1,2$) (\ref{aux_def}) are combinations of the
{\em independent} Airy functions and their derivatives,
\begin{equation}\label{aux_Airy}
\begin{split}
&\hat\psi_n^{(j)}(\xi)=
\frac{1}{h_n}p_n(\partial_{\tau})\int_{\Gamma_j}
e^{-\frac{1}{3}\lambda^3+\tau\lambda}\,d\lambda\Bigr|_{\tau=t\xi-x},
\\
&\hat\phi_m^{(j)}(\lambda)=
\frac{1}{h_m}q_m(\partial_{\tau})\int_{\Gamma_j}
e^{-\frac{1}{3}\xi^3+\tau\xi}\,d\xi\bigr|_{\tau=t\lambda-y},
\end{split}
\end{equation}
we construct the particular piece-wise holomorphic solutions of the
system of the $3\times3$ matrix equations (\ref{matrix_eqs_dual_Psi}) 
and (\ref{matrix_eqs_dual_Phi}),
\begin{equation}\label{dual_Psi_sol}
\hat\Psi_n(\xi)=
\begin{pmatrix}
\hat\psi_n^{(1)}(\xi)&
\hat\psi_n^{(2)}(\xi)&
F_n(\xi)\\
\hat\psi_{n-1}^{(1)}(\xi)&
\hat\psi_{n-1}^{(2)}(\xi)&
F_{n-1}(\xi)\\
\hat\psi_{n-2}^{(1)}(\xi)&
\hat\psi_{n-2}^{(2)}(\xi)&
F_{n-2}(\xi)
\end{pmatrix},\quad
n\geq4,
\end{equation}
and
\begin{equation}\label{dual_Phi_sol}
\hat\Phi_m(\lambda)=
\begin{pmatrix}
\hat\phi_m^{(1)}(\lambda)&
\hat\phi_m^{(2)}(\lambda)&
G_m(\lambda)\\
\hat\phi_{m-1}^{(1)}(\lambda)&
\hat\phi_{m-1}^{(2)}(\lambda)&
G_{m-1}(\lambda)\\
\hat\phi_{m-2}^{(1)}(\lambda)&
\hat\phi_{m-2}^{(2)}(\lambda)&
G_{m-2}(\lambda)
\end{pmatrix},\quad
m\geq4.
\end{equation}
The jump property of the Cauchy integral yields the relations
\begin{equation}\label{F_n_jump}
\begin{split}
&F_n^+(\xi)-F_n^-(\xi)=
(g_j^{(2)}-g_{j+1}^{(2)})\hat\psi_n(\xi),\quad
\arg\xi=\frac{2\pi}{3}(j-1),\quad
j=0,1,2,
\\
&G_m^+(\lambda)-G_m^-(\lambda)=
(g_j^{(1)}-g_{j+1}^{(1)})\hat\phi_m(\lambda),\quad
\arg\lambda=\frac{2\pi}{3}(j-1),\quad
j=0,1,2,
\end{split}
\end{equation}
where $g_3^{(i)}\equiv g_0^{(i)}$. Thus we find that the matrix
functions $\hat\Psi_n(\xi)$ and $\hat\Phi_m(\lambda)$ have the
following jumps across the rays 
$\ell_j=\bigl\{\xi\in{\Bbb C}\colon\
\arg\xi=\frac{2\pi}{3}(j-1)\bigr\}$, $j=0,1,2$, oriented towards infinity,
\begin{equation}\label{hat_Psi_n_jump}
\begin{split}
&\hat\Psi_n^+(\xi)=\hat\Psi_n^-(\xi)
\begin{pmatrix}
1&0&(g_j^{(2)}-g_{j+1}^{(2)})(g_1^{(1)}-g_0^{(1)})\\
0&1&(g_j^{(2)}-g_{j+1}^{(2)})(g_2^{(1)}-g_0^{(1)})\\
0&0&1
\end{pmatrix},\quad
\xi\in\ell_j,\quad
j=0,1,2,
\\
&\hat\Phi_n^+(\lambda)=\hat\Phi_n^-(\lambda)
\begin{pmatrix}
1&0&(g_j^{(1)}-g_{j+1}^{(1)})(g_1^{(2)}-g_0^{(2)})\\
0&1&(g_j^{(1)}-g_{j+1}^{(1)})(g_2^{(2)}-g_0^{(2)})\\
0&0&1
\end{pmatrix},\quad
\lambda\in\ell_j,\quad
j=0,1,2.
\end{split}
\end{equation}

Using the well known asymptotics of the Airy integrals in the complex
domain,
\begin{equation}\label{y1_as}
\begin{split}
&\int_{\Gamma_0}
e^{-\frac{1}{3}\lambda^3+\tau\lambda}\,d\lambda=
-i\sqrt\pi\,\tau^{-1/4}e^{-\frac{2}{3}\tau^{3/2}}\bigl(
1+{\cal O}(\tau^{-3/2})\bigr),\quad
\arg\tau\in(-\tfrac{2\pi}{3},\tfrac{2\pi}{3}),
\\
&\int_{\Gamma_1}
e^{-\frac{1}{3}\lambda^3+\tau\lambda}\,d\lambda=
i\sqrt\pi\,\tau^{-1/4}e^{-\frac{2}{3}\tau^{3/2}}\bigl(
1+{\cal O}(\tau^{-3/2})\bigr),\quad
\arg\tau\in(\tfrac{2\pi}{3},2\pi),
\\
&\int_{\Gamma_2}
e^{-\frac{1}{3}\lambda^3+\tau\lambda}\,d\lambda=
-\sqrt\pi\,\tau^{-1/4}e^{\frac{2}{3}\tau^{3/2}}\bigl(
1+{\cal O}(\tau^{-3/2})\bigr),\quad
\arg\tau\in(0,\tfrac{4\pi}{3}),
\end{split}
\end{equation}
the asymptotics of the Cauchy integrals for (\ref{F_n_def}),
\begin{equation}\label{F_n_as}
\begin{split}
&F_n(\xi)=-\frac{h_n}{2\pi i}\xi^{-n-1}e^{W(\xi)}
\bigl(1+{\cal O}(\xi^{-1})\bigr),
\\
&G_m(\lambda)=-\frac{h_m}{2\pi i}\lambda^{-m-1}e^{V(\lambda)}
\bigl(1+{\cal O}(\lambda^{-1})\bigr),
\end{split}
\end{equation}
we construct {\bf the RH problem for the dual functions $\hat\psi_n(\xi)$.} 

\begin{RHP}\label{RHP_dual_psi}
Find a piece-wise holomorphic $3\times3$ matrix function 
$\hat\Psi_n^{RH}(\xi)$ with the following properties:

1. As $\xi\to\infty$,
\begin{equation}\label{hat_Psi_n_as_RH}
\begin{split}
\hat\Psi_n^{RH}(\xi)\to
&\begin{pmatrix}
\frac{\sqrt\pi}{h_n}(t\xi)^{\frac{n}{2}-\frac{1}{4}}&
\frac{\sqrt\pi}{h_n}(-1)^n(t\xi)^{\frac{n}{2}-\frac{1}{4}}&
-\frac{h_n}{2\pi i}\xi^{-n-1}\\
\frac{\sqrt\pi}{h_{n-1}}(t\xi)^{\frac{n-1}{2}-\frac{1}{4}}&
\frac{\sqrt\pi}{h_{n-1}}(-1)^{n-1}(t\xi)^{\frac{n-1}{2}-\frac{1}{4}}&
-\frac{h_{n-1}}{2\pi i}\xi^{-n}\\
\frac{\sqrt\pi}{h_{n-2}}(t\xi)^{\frac{n-2}{2}-\frac{1}{4}}&
\frac{\sqrt\pi}{h_{n-2}}(-1)^{n-2}(t\xi)^{\frac{n-2}{2}-\frac{1}{4}}&
-\frac{h_{n-2}}{2\pi i}\xi^{-n+1}
\end{pmatrix}\times
\\
&\times\begin{pmatrix}
e^{\frac{2}{3}(t\xi)^{3/2}-x(t\xi)^{1/2}}&0&\\
0&e^{-\frac{2}{3}(t\xi)^{3/2}+x(t\xi)^{1/2}}&0\\
0&0&e^{\frac{1}{3}\xi^3+y\xi}
\end{pmatrix};
\end{split}
\end{equation}

2. Across the rays $\arg\xi=\frac{2\pi}{3}(j-1)$, $j=1,2,3$, oriented
towards infinity, $\hat\Psi_n^{RH}(\xi)$ has the jumps
\begin{equation}\label{hat_Psi_RH_jumps}
\hat\Psi_n^{RH+}(\xi)=\hat\Psi_n^{RH-}(\xi)\hat S_j,\quad
\arg\xi=\frac{2\pi}{3}(j-1),
\end{equation}
where plus and minus indicate the limiting values of
$\Psi_n^{RH}(\xi)$ on the jump contour from the left and from the right,
respectively, and
$$
\hat S_1=\begin{pmatrix}
1&0&(g_1^{(2)}-g_{2}^{(2)})(g_1^{(1)}-g_2^{(1)})\\
-i&1&i(g_1^{(2)}-g_{2}^{(2)})(g_2^{(1)}-g_0^{(1)})\\
0&0&1
\end{pmatrix},
$$
$$
\hat S_2=\begin{pmatrix}
1&-i&(g_2^{(2)}-g_0^{(2)})(g_1^{(1)}-g_2^{(1)})\\
0&1&i(g_2^{(2)}-g_0^{(2)})(g_1^{(1)}-g_0^{(1)})\\
0&0&1
\end{pmatrix},
$$
$$
\hat S_3=\begin{pmatrix}
1&0&(g_0^{(2)}-g_{1}^{(2)})(g_0^{(1)}-g_2^{(1)})\\
-i&1&i(g_0^{(2)}-g_{1}^{(2)})(g_1^{(1)}-g_0^{(1)})\\
0&0&1
\end{pmatrix},
$$
and across the ray $\arg\xi=-\frac{\pi}{3}$ oriented towards infinity,
the following jump condition holds,
\begin{equation}\label{hat_Psi_branch_jump}
\hat\Psi_n^{RH+}(\xi)=\hat\Psi_n^{RH-}(\xi)\hat\Sigma,\quad
\arg\xi=-\frac{\pi}{3},\quad
\hat\Sigma=\begin{pmatrix}
0&i&0\\
i&0&0\\
0&0&1
\end{pmatrix}.
\end{equation}
\end{RHP}

The dual functions $\hat\psi_n(\xi)$, $\hat\psi_{n-1}(\xi)$ and
$\hat\psi_{n-2}(\xi)$ form the vector $\hat\Psi_n(\xi)$ related to the 
solution of the RH problem~\ref{RHP_dual_psi} by the following equations,
\begin{equation}\label{dual_psi_sol}
\hat\Psi_n(\xi)=\begin{cases}
\hat\Psi_n^{RH}(\xi)R^{(1)},\quad&
\arg\xi\in(-\frac{\pi}{3},0),\\
\hat\Psi_n^{RH}(\xi)
\hat S_1^{-1}R^{(1)},\quad&
\arg\xi\in(0,\frac{2\pi}{3}),\\
\hat\Psi_n^{RH}(\xi)
\hat S_2^{-1}\hat S_1^{-1}R^{(1)},\quad&
\arg\xi\in(\frac{2\pi}{3},\frac{4\pi}{3}),\\
\hat\Psi_n^{RH}(\xi)
\hat S_3^{-1}\hat S_2^{-1}\hat S_1^{-1}R^{(1)},\quad&
\arg\xi\in(\frac{4\pi}{3},\frac{5\pi}{3}),
\end{cases}
\end{equation}
where
$$
R^{(1)}=\begin{pmatrix}
g_1^{(1)}-g_2^{(1)}\\
i(g_2^{(1)}-g_0^{(1)})\\
0
\end{pmatrix}.
$$

\begin{RHP}\label{RHP_dual_phi}
Find a piece-wise holomorphic $3\times3$ matrix function 
$\hat\Phi_m^{RH}(\lambda)$ with the following properties:

1. As $\lambda\to\infty$,
\begin{equation}\label{hat_Phi_m_as_RH}
\begin{split}
\hat\Phi_m^{RH}(\lambda)\to
&\begin{pmatrix}
\frac{\sqrt\pi}{h_m}(t\lambda)^{\frac{m}{2}-\frac{1}{4}}&
\frac{\sqrt\pi}{h_m}(-1)^m(t\lambda)^{\frac{m}{2}-\frac{1}{4}}&
-\frac{h_m}{2\pi i}\lambda^{-m-1}\\
\frac{\sqrt\pi}{h_{m-1}}(t\lambda)^{\frac{m-1}{2}-\frac{1}{4}}&
\frac{\sqrt\pi}{h_{m-1}}(-1)^{m-1}(t\lambda)^{\frac{m-1}{2}-\frac{1}{4}}&
-\frac{h_{m-1}}{2\pi i}\lambda^{-m}\\
\frac{\sqrt\pi}{h_{m-2}}(t\lambda)^{\frac{m-2}{2}-\frac{1}{4}}&
\frac{\sqrt\pi}{h_{m-2}}(-1)^{m-2}(t\lambda)^{\frac{m-2}{2}-\frac{1}{4}}&
-\frac{h_{m-2}}{2\pi i}\lambda^{-m+1}
\end{pmatrix}\times
\\
&\times\begin{pmatrix}
e^{\frac{2}{3}(t\lambda)^{3/2}-y(t\lambda)^{1/2}}&0&\\
0&e^{-\frac{2}{3}(t\lambda)^{3/2}+y(t\lambda)^{1/2}}&0\\
0&0&e^{\frac{1}{3}\lambda^3+x\lambda}
\end{pmatrix};
\end{split}
\end{equation}

2. Across the rays $\arg\lambda=\frac{2\pi}{3}(j-1)$, $j=0,1,2$, oriented
towards infinity, $\hat\Phi_m^{RH}(\lambda)$ has the jumps
\begin{equation}\label{hat_Phi_RH_jumps}
\hat\Phi_m^{RH+}(\lambda)=\hat\Phi_m^{RH-}(\lambda)\hat T_j,\quad
\arg\lambda=\frac{2\pi}{3}(j-1),
\end{equation}
where plus and minus indicate the left and right limits of
$\Phi_m^{RH}(\lambda)$ on the jump contour, and
$$
\hat T_1=\begin{pmatrix}
1&0&(g_1^{(1)}-g_{2}^{(1)})(g_1^{(2)}-g_2^{(2)})\\
-i&1&i(g_1^{(1)}-g_{2}^{(1)})(g_2^{(2)}-g_0^{(2)})\\
0&0&1
\end{pmatrix},
$$
$$
\hat T_2=\begin{pmatrix}
1&-i&(g_2^{(1)}-g_0^{(1)})(g_1^{(2)}-g_2^{(2)})\\
0&1&i(g_2^{(1)}-g_0^{(1)})(g_1^{(2)}-g_0^{(2)})\\
0&0&1
\end{pmatrix},
$$
$$
\hat T_3=\begin{pmatrix}
1&0&(g_0^{(1)}-g_{1}^{(1)})(g_0^{(2)}-g_2^{(2)})\\
-i&1&i(g_0^{(1)}-g_{1}^{(1)})(g_1^{(2)}-g_0^{(2)})\\
0&0&1
\end{pmatrix},
$$
and across the ray $\arg\lambda=-\frac{\pi}{3}$ oriented towards
infinity, the following jump condition holds,
\begin{equation}\label{hat_Phi_branch_jump}
\hat\Phi_m^{RH+}(\lambda)=\hat\Phi_n^{RH-}(\lambda)\hat\Sigma,\quad
\arg\lambda=-\frac{\pi}{3},\quad
\hat\Sigma=\begin{pmatrix}
0&i&0\\
i&0&0\\
0&0&1
\end{pmatrix}.
\end{equation}
\end{RHP}

The dual functions $\hat\phi_m(\lambda)$, $\hat\phi_{m-1}(\lambda)$ and
$\hat\phi_{m-2}(\lambda)$ form the vector $\hat\Phi_m(\lambda)$ related 
to the solution of the RH problem~\ref{RHP_dual_phi} by the equations
\begin{equation}\label{dual_phi_sol}
\hat\Phi_m(\lambda)=\begin{cases}
\hat\Phi_m^{RH}(\lambda)R^{(2)},\quad&
\arg\lambda\in(-\frac{\pi}{3},0),\\
\hat\Phi_m^{RH}(\lambda)
\hat T_1^{-1}R^{(2)},\quad&
\arg\lambda\in(0,\frac{2\pi}{3}),\\
\hat\Phi_m^{RH}(\lambda)
\hat T_2^{-1}\hat T_1^{-1}R^{(2)},\quad&
\arg\lambda\in(\frac{2\pi}{3},\frac{4\pi}{3}),\\
\hat\Phi_m^{RH}(\lambda)
\hat T_3^{-1}\hat T_2^{-1}\hat T_1^{-1}R^{(2)},\quad&
\arg\lambda\in(\frac{4\pi}{3},\frac{5\pi}{3}),
\end{cases}
\end{equation}
where
$$
R^{(2)}=\begin{pmatrix}
g_1^{(2)}-g_2^{(2)}\\
i(g_2^{(2)}-g_0^{(2)})\\
0
\end{pmatrix}.
$$

\section{The particular solutions of the matrix equations and
the Riemann-Hilbert problem for the wave functions}
\label{wave_RH}

The method of construction of the Riemann-Hilbert problem for the wave
functions $\psi_n(\lambda)$ and $\phi_m(\xi)$ is more involved because
of less elementary structure of their integral representations in
compare to those for dual functions. 

Let $\tilde\ell_0^{(j)}$ be an oriented contour connecting a finite
point $\xi_0$ (or $\lambda_0$) with infinity within the sector of the
exponential decay of the function $\hat\psi_n^{(j)}(\xi)$
(\ref{aux_Airy}) (resp., of $\hat\phi_m^{(j)}(\lambda)$). Namely, let 
$\tilde\ell_0^{(j)}$ be asymptotic to the ray
\begin{equation}\label{tilde_ell_def}
\tilde\ell_0^{(j)}\sim[0,e^{-\frac{2\pi}{3}j}\infty).
\end{equation}

Let $\tilde\Gamma_j$ be an infinite oriented contour asymptotic to the rays
\begin{equation}\label{tilde_Gamma_def}
\tilde\Gamma_j\sim(e^{i\frac{2\pi}{3}(j-\frac{3}{2})}\infty,0]\cup
[0,e^{i\frac{2\pi}{3}(j-\frac{1}{2})}\infty),\quad 
j=0,1,2,
\end{equation}
located within the sector (\ref{tilde_Gamma_def}) and intersecting the
ray
\begin{equation}\label{ell_def}
\ell_j=[0,e^{i\frac{2\pi}{3}(j-1)}\infty),
\end{equation}
so that $\tilde\Gamma_j\cap\ell_j=\{\xi_j\}$ (resp.,
$\tilde\Gamma_j\cap\ell_j=\{\lambda_j\}$). 

The ``bricks'' which we use to build up the appropriate integral
representations are the ``inverse'' Laplace-Fourier transforms 
\begin{equation}\label{bricks_psi}
\tilde\psi_n^{(j)}(\lambda)=
\frac{t}{\pi i}\int_{\tilde\ell_0^{(j)}}
e^{-t\lambda\xi}\hat\psi_n^{(j)}(\xi)\,d\xi,\quad
\tilde F_n(\lambda)=
\frac{t}{\pi i}\int_{\tilde\Gamma_j}F_n(\xi)e^{-t\lambda\xi}\,d\xi,
\end{equation}
and
\begin{equation}\label{bricks_phi}
\tilde\phi_m^{(j)}(\xi)=
\frac{t}{\pi i}\int_{\tilde\ell_0^{(j)}}
e^{-t\lambda\xi}\hat\phi_m^{(j)}(\lambda)\,d\lambda,\quad
\tilde G_m(\xi)=
\frac{t}{\pi i}\int_{\tilde\Gamma_j}G_m(\lambda)e^{-t\lambda\xi}\,d\lambda,
\end{equation}
The functions (\ref{bricks_psi}) satisfy the differential equations in
$\lambda$, $x$ and $y$ in (\ref{eqs_psi}) but not the recursion
relation and the differential equation in $t$ because of the
appearance of the inappropriate off-integral terms after integration
by parts. The similar observation holds true for (\ref{bricks_phi}).
More in details,
\begin{equation}\label{tilde_psi_off}
\lambda\tilde\psi_n^{(k)}(\lambda)=\frac{1}{\pi i}e^{-t\lambda\xi_0}
\hat\psi_n^{(k)}(\xi_0)+\mbox{appropriate terms},
\end{equation}
\begin{equation}\label{tilde_F_off}
\begin{split}
\lambda\tilde F_n(\lambda)&=
\frac{1}{\pi i}e^{-t\lambda\xi_j}\bigl(F_n^+(\xi_j)-F_n^-(\xi_j)\bigr)+
\mbox{appropriate terms}=
\\
&=\frac{1}{\pi i}e^{-t\lambda\xi_j}
(g_j^{(2)}-g_{j+1}^{(2)})\sum_{k=0}^2g_k^{(1)}\hat\psi_n^{(k)}(\xi_j)+
\mbox{appropriate terms}.
\end{split}
\end{equation}
In second line of (\ref{tilde_F_off}), we have used the jump condition
(\ref{F_n_jump}) and the definition (\ref{dual_via_aux}). Combining
(\ref{tilde_psi_off}) at $\xi_0=\xi_j$ and (\ref{tilde_F_off}), it is
possible to eliminate the off-integral terms and find such a
combination $\tilde F_n^{(j)}(\lambda)$ of
$\tilde\psi_n^{(k)}(\lambda)$ and $\tilde F_n(\lambda)$ which
satisfies the system (\ref{eqs_psi}), 
\begin{equation}\label{tilde_F_n^j}
\tilde F_n^{(j)}(\lambda)=\tilde F_n(\lambda)
-(g_j^{(2)}-g_{j+1}^{(2)})\sum_{k=0}^2g_k^{(1)}
\tilde\psi_n^{(k)}(\lambda).
\end{equation}

Using identity $\sum_j\hat\psi^{(j)}(\xi)=0$, it is possible to
eliminate one of the values $\hat\psi_n^{(k)}(\xi_j)$ from
(\ref{tilde_F_off}) and therefore the respective term in
(\ref{tilde_F_n^j}). By technical reasons, we prefer to use the latter
possibility and introduce the following solutions of (\ref{eqs_psi}):
\begin{equation}\label{tilde_F_n^0_def}
\tilde F_n^{(0)}(\lambda)=
\tilde F_n(\lambda)
-(g_0^{(2)}-g_1^{(2)})(g_0^{(1)}-g_1^{(1)})\tilde\psi_n^{(0)}(\lambda)
-(g_0^{(2)}-g_1^{(2)})(g_2^{(1)}-g_1^{(1)})\tilde\psi_n^{(2)}(\lambda),
\end{equation}
\begin{equation}\label{tilde_F_n^1_def}
\tilde F_n^{(1)}(\lambda)=
\tilde F_n(\lambda)
-(g_1^{(2)}-g_2^{(2)})(g_1^{(1)}-g_0^{(1)})\tilde\psi_n^{(1)}(\lambda)
-(g_1^{(2)}-g_2^{(2)})(g_2^{(1)}-g_0^{(1)})\tilde\psi_n^{(2)}(\lambda),
\end{equation}
\begin{equation}\label{tilde_F_n^2_def}
\tilde F_n^{(2)}(\lambda)=
\tilde F_n(\lambda)
-(g_2^{(2)}-g_0^{(2)})(g_0^{(1)}-g_2^{(1)})\tilde\psi_n^{(0)}(\lambda)
-(g_2^{(2)}-g_0^{(2)})(g_1^{(1)}-g_2^{(1)})\tilde\psi_n^{(1)}(\lambda),
\end{equation}
In the very same way,
\begin{equation}\label{tilde_G_m^0_def}
\tilde G_m^{(0)}(\xi)=\tilde G_m(\xi)
-(g_0^{(1)}-g_1^{(1)})(g_0^{(2)}-g_1^{(2)})\tilde\phi_m^{(0)}(\xi)
-(g_0^{(1)}-g_1^{(1)})(g_2^{(2)}-g_1^{(2)})\tilde\phi_m^{(2)}(\xi),
\end{equation}
\begin{equation}\label{tilde_G_m^1_def}
\tilde G_m^{(1)}(\xi)=
\tilde G_m(\xi)
-(g_1^{(1)}-g_2^{(1)})(g_1^{(2)}-g_0^{(2)})\tilde\phi_m^{(1)}(\xi)
-(g_1^{(1)}-g_2^{(1)})(g_2^{(2)}-g_0^{(2)})\tilde\phi_m^{(2)}(\xi),
\end{equation}
\begin{equation}\label{tilde_G_m^2_def}
\tilde G_m^{(2)}(\xi)=
\tilde G_m(\xi)
-(g_2^{(1)}-g_0^{(1)})(g_0^{(2)}-g_2^{(2)})\tilde\phi_m^{(0)}(\xi)
-(g_2^{(1)}-g_0^{(1)})(g_1^{(2)}-g_2^{(2)})\tilde\phi_m^{(1)}(\xi).
\end{equation}

Define the $3\times3$ matrix functions
\begin{equation}\label{Psi_sol}
\Psi_n(\lambda)=\begin{pmatrix}
\psi_n(\lambda)&\tilde F_n^{(0)}(\lambda)&\tilde F_n^{(1)}(\lambda)\\
\psi_{n-1}(\lambda)&\tilde F_{n-1}^{(0)}(\lambda)&
\tilde F_{n-1}^{(1)}(\lambda)\\
\psi_{n-2}(\lambda)&\tilde F_{n-2}^{(0)}(\lambda)&
\tilde F_{n-2}^{(1)}(\lambda)
\end{pmatrix},\quad
n\geq4,
\end{equation}
and
\begin{equation}\label{Phi_sol}
\Phi_m(\xi)=\begin{pmatrix}
\phi_m(\xi)&\tilde G_m^{(0)}(\xi)&\tilde G_m^{(1)}(\xi)\\
\phi_{m-1}(\xi)&\tilde G_{m-1}^{(0)}(\xi)&\tilde G_{m-1}^{(1)}(\xi)\\
\phi_{m-2}(\xi)&\tilde G_{m-2}^{(0)}(\xi)&\tilde G_{m-2}^{(1)}(\xi)
\end{pmatrix},\quad
m\geq4.
\end{equation}

The asymptotics at infinity of $\psi_n(\lambda)$ and $\phi_m(\xi)$ is
elementary,
\begin{equation}\label{psi_n_phi_m_as}
\psi_n(\lambda)=
\frac{\lambda^n}{h_n}e^{-\frac{1}{3}\lambda^3-x\lambda}
\bigl(1+{\cal O}(\lambda^{-1})\bigr),\quad
\phi_m(\xi)=
\frac{\xi^m}{h_m}e^{-\frac{1}{3}\xi^3-y\xi}
\bigl(1+{\cal O}(\xi^{-1})\bigr).
\end{equation}
The asymptotics of $\tilde F_n^{(j)}(\lambda)$ and 
$\tilde G_m^{(j)}(\xi)$ can be found using the conventional steepest
descent method,
\begin{multline}\label{tilde_Fn^1_as_1}
\tilde F_n^{(1)}(\lambda)
=\frac{ith_n}{2\pi^{3/2}}
(t\lambda)^{-\frac{n+1}{2}-\frac{1}{4}}
e^{-\frac{2}{3}(t\lambda)^{3/2}+y(t\lambda)^{1/2}}
\bigl(1+{\cal O}(\lambda^{-1/2})\bigr)+
\\
+2(g_1^{(2)}-g_2^{(2)})(g_1^{(1)}-g_0^{(1)})
\frac{\lambda^n}{h_n}e^{-\frac{1}{3}\lambda^3-x\lambda}
\bigl(1+{\cal O}(\lambda^{-1})\bigr),
\quad
\arg\lambda\in(-\tfrac{2\pi}{3},0),
\end{multline}
\begin{multline}\label{tilde_Fn^1_as_2}
\tilde F_n^{(1)}(\lambda)
=\frac{ith_n}{2\pi^{3/2}}
(t\lambda)^{-\frac{n+1}{2}-\frac{1}{4}}
e^{-\frac{2}{3}(t\lambda)^{3/2}+y(t\lambda)^{1/2}}
\bigl(1+{\cal O}(\lambda^{-1/2})\bigr)+
\\
+2(g_1^{(2)}-g_2^{(2)})(g_2^{(1)}-g_0^{(1)})
\frac{\lambda^n}{h_n}e^{-\frac{1}{3}\lambda^3-x\lambda}
\bigl(1+{\cal O}(\lambda^{-1})\bigr),
\quad
\arg\lambda\in(0,\tfrac{2\pi}{3}),
\end{multline}
\begin{multline}\label{tilde_Fn^0_as_1}
\tilde F_n^{(0)}(\lambda)=
-\frac{th_n}{2\pi^{3/2}}(-1)^n
(t\lambda)^{-\frac{n+1}{2}-\frac{1}{4}}
e^{\frac{2}{3}(t\lambda)^{3/2}-y(t\lambda)^{1/2}}
\bigl(1+{\cal O}(\lambda^{-1/2})\bigr)+
\\
+2(g_0^{(2)}-g_1^{(2)})(g_2^{(1)}-g_1^{(1)})
\frac{\lambda^n}{h_n}e^{-\frac{1}{3}\lambda^3-x\lambda}
\bigl(1+{\cal O}(\lambda^{-1})\bigr),
\quad
\arg\lambda\in(0,\tfrac{2\pi}{3}),
\end{multline}
\begin{multline}\label{tilde_Fn^0_as_2}
\tilde F_n^{(0)}(\lambda)=
-\frac{th_n}{2\pi^{3/2}}(-1)^n
(t\lambda)^{-\frac{n+1}{2}-\frac{1}{4}}
e^{\frac{2}{3}(t\lambda)^{3/2}-y(t\lambda)^{1/2}}
\bigl(1+{\cal O}(\lambda^{-1/2})\bigr)+
\\
+2(g_0^{(2)}-g_1^{(2)})(g_0^{(1)}-g_1^{(1)})
\frac{\lambda^n}{h_n}e^{-\frac{1}{3}\lambda^3-x\lambda}
\bigl(1+{\cal O}(\lambda^{-1})\bigr),
\quad
\arg\lambda\in(\tfrac{2\pi}{3},\tfrac{4\pi}{3}),
\end{multline}
and
\begin{multline}\label{tilde_Fn^2_as_1}
\tilde F_n^{(2)}(\lambda)=
-\frac{ith_n}{2\pi^{3/2}}
(t\lambda)^{-\frac{n+1}{2}-\frac{1}{4}}
e^{-\frac{2}{3}(t\lambda)^{3/2}+y(t\lambda)^{1/2}}
\bigl(1+{\cal O}(\lambda^{-1/2})\bigr)+
\\
+2(g_2^{(2)}-g_0^{(2)})(g_0^{(1)}-g_2^{(1)})
\frac{\lambda^n}{h_n}e^{-\frac{1}{3}\lambda^3-x\lambda}
\bigl(1+{\cal O}(\lambda^{-1})\bigr),
\quad
\arg\lambda\in(\tfrac{2\pi}{3},\tfrac{4\pi}{3}),
\end{multline}
\begin{multline}\label{tilde_Fn^2_as_2}
\tilde F_n^{(2)}(\lambda)=
-\frac{ith_n}{2\pi^{3/2}}
(t\lambda)^{-\frac{n+1}{2}-\frac{1}{4}}
e^{-\frac{2}{3}(t\lambda)^{3/2}+y(t\lambda)^{1/2}}
\bigl(1+{\cal O}(\lambda^{-1/2})\bigr)+
\\
+2(g_2^{(2)}-g_0^{(2)})(g_1^{(1)}-g_2^{(1)})
\frac{\lambda^n}{h_n}e^{-\frac{1}{3}\lambda^3-x\lambda}
\bigl(1+{\cal O}(\lambda^{-1})\bigr),
\quad
\arg\lambda\in(\tfrac{4\pi}{3},2\pi).
\end{multline}

For the asymptotics of $\tilde G_m(\xi)$, we have
\begin{multline}\label{tilde_Gm^1_as_1}
\tilde G_m^{(1)}(\xi)
=\frac{ith_m}{2\pi^{3/2}}
(t\xi)^{-\frac{m+1}{2}-\frac{1}{4}}
e^{-\frac{2}{3}(t\xi)^{3/2}+x(t\xi)^{1/2}}
\bigl(1+{\cal O}(\xi^{-1/2})\bigr)+
\\
+2(g_1^{(1)}-g_2^{(1)})(g_1^{(2)}-g_0^{(2)})
\frac{\xi^m}{h_m}e^{-\frac{1}{3}\xi^3-y\xi}
\bigl(1+{\cal O}(\xi^{-1})\bigr),
\quad
\arg\xi\in(-\tfrac{2\pi}{3},0)
\end{multline}
\begin{multline}\label{tilde_Gm^1_as_2}
\tilde G_m^{(1)}(\xi)
=\frac{ith_m}{2\pi^{3/2}}
(t\xi)^{-\frac{m+1}{2}-\frac{1}{4}}
e^{-\frac{2}{3}(t\xi)^{3/2}+x(t\xi)^{1/2}}
\bigl(1+{\cal O}(\xi^{-1/2})\bigr)+
\\
+2(g_1^{(1)}-g_2^{(1)})(g_2^{(2)}-g_0^{(2)})
\frac{\xi^m}{h_m}e^{-\frac{1}{3}\xi^3-y\xi}
\bigl(1+{\cal O}(\xi^{-1})\bigr),
\quad
\arg\xi\in(0,\tfrac{2\pi}{3})
\end{multline}
\begin{multline}\label{tilde_Gm^0_as_1}
\tilde G_m^{(0)}(\xi)=
-\frac{th_m}{2\pi^{3/2}}(-1)^m
(t\xi)^{-\frac{m+1}{2}-\frac{1}{4}}
e^{\frac{2}{3}(t\xi)^{3/2}-x(t\xi)^{1/2}}
\bigl(1+{\cal O}(\xi^{-1/2})\bigr)+
\\
+2(g_0^{(1)}-g_1^{(1)})(g_2^{(2)}-g_1^{(2)})
\frac{\xi^m}{h_m}e^{-\frac{1}{3}\xi^3-y\xi}
\bigl(1+{\cal O}(\xi^{-1})\bigr),
\quad
\arg\xi\in(0,\tfrac{2\pi}{3}),
\end{multline}
\begin{multline}\label{tilde_Gm^0_as_2}
\tilde G_m^{(0)}(\xi)=
-\frac{th_m}{2\pi^{3/2}}(-1)^m
(t\xi)^{-\frac{m+1}{2}-\frac{1}{4}}
e^{\frac{2}{3}(t\xi)^{3/2}-x(t\xi)^{1/2}}
\bigl(1+{\cal O}(\xi^{-1/2})\bigr)+
\\
+2(g_0^{(1)}-g_1^{(1)})(g_0^{(2)}-g_1^{(2)})
\frac{\xi^m}{h_m}e^{-\frac{1}{3}\xi^3-y\xi}
\bigl(1+{\cal O}(\xi^{-1})\bigr),
\quad
\arg\xi\in(\tfrac{2\pi}{3},\tfrac{4\pi}{3}),
\end{multline}
and
\begin{multline}\label{tilde_Gm^2_as_1}
\tilde G_m^{(2)}(\xi)=
-\frac{ith_m}{2\pi^{3/2}}
(t\xi)^{-\frac{m+1}{2}-\frac{1}{4}}
e^{-\frac{2}{3}(t\xi)^{3/2}+x(t\xi)^{1/2}}
\bigl(1+{\cal O}(\xi^{-1/2})\bigr)+
\\
+2(g_2^{(1)}-g_0^{(1)})(g_0^{(2)}-g_2^{(2)})
\frac{\xi^m}{h_m}e^{-\frac{1}{3}\xi^3-y\xi}
\bigl(1+{\cal O}(\xi^{-1})\bigr),
\quad
\arg\xi\in(\tfrac{2\pi}{3},\tfrac{4\pi}{3}),
\end{multline}
\begin{multline}\label{tilde_Gm^2_as_2}
\tilde G_m^{(2)}(\xi)=
-\frac{ith_m}{2\pi^{3/2}}
(t\xi)^{-\frac{m+1}{2}-\frac{1}{4}}
e^{-\frac{2}{3}(t\xi)^{3/2}+x(t\xi)^{1/2}}
\bigl(1+{\cal O}(\xi^{-1/2})\bigr)+
\\
+2(g_2^{(1)}-g_0^{(1)})(g_1^{(2)}-g_2^{(2)})
\frac{\xi^m}{h_m}e^{-\frac{1}{3}\xi^3-y\xi}
\bigl(1+{\cal O}(\xi^{-1})\bigr),
\quad
\arg\xi\in(\tfrac{4\pi}{3},2\pi).
\end{multline}

Using the above asymptotics and the linear constraints for
$\tilde F_n^{(j)}$, $\tilde G_m^{(j)}$,
\begin{multline}\label{global_F_n}
\tilde F_n^{(0)}(\lambda)+\tilde F_n^{(1)}(\lambda)
+\tilde F_n^{(2)}(\lambda)=2g_F\psi_n(\lambda),
\\
g_F=g_0^{(2)}(g_2^{(1)}-g_1^{(1)})
+g_1^{(2)}(g_1^{(1)}-g_0^{(1)})
+g_2^{(2)}(g_0^{(1)}-g_2^{(1)}),
\end{multline}
and
\begin{multline}\label{global_G_m}
\tilde G_m^{(0)}(\xi)+\tilde G_m^{(1)}(\xi)
+\tilde G_m^{(2)}(\xi)
=2g_G\phi_m(\xi),
\\
g_G=g_0^{(1)}(g_2^{(2)}-g_1^{(2)})
+g_1^{(1)}(g_1^{(2)}-g_0^{(2)})
+g_2^{(1)}(g_0^{(2)}-g_2^{(2)}),
\end{multline}
we find the RH problems for our bi-orthogonal polynomials.

\begin{RHP}\label{RHP_psi}
Find a piece-wise holomorphic $3\times3$ matrix function 
$\Psi_n^{RH}(\lambda)$ with the following properties:

1. As $\lambda\to\infty$,
\begin{equation}\label{Psi_n_as_RH}
\begin{split}
\Psi_n^{RH}(\lambda)\to&
\begin{pmatrix}
\frac{\lambda^n}{h_n}
&-\frac{th_n}{2\pi^{3/2}}
(-1)^n(t\lambda)^{-\frac{n+1}{2}-\frac{1}{4}}
&\frac{ith_n}{2\pi^{3/2}}
(t\lambda)^{-\frac{n+1}{2}-\frac{1}{4}}
\\
\frac{\lambda^{n-1}}{h_{n-1}}
&-\frac{th_{n-1}}{2\pi^{3/2}}
(-1)^{n-1}
(t\lambda)^{-\frac{n}{2}-\frac{1}{4}}
&\frac{ith_{n-1}}{2\pi^{3/2}}
(t\lambda)^{-\frac{n}{2}-\frac{1}{4}}
\\
\frac{\lambda^{n-2}}{h_{n-2}}
&-\frac{th_{n-2}}{2\pi^{3/2}}
(-1)^{n-2}
(t\lambda)^{-\frac{n-1}{2}-\frac{1}{4}}
&\frac{ith_{n-2}}{2\pi^{3/2}}
(t\lambda)^{-\frac{n-1}{2}-\frac{1}{4}}
\end{pmatrix}\times
\\
&\times
\begin{pmatrix}
e^{-\frac{1}{3}\lambda^3-x\lambda}&0&0\\
0&e^{\frac{2}{3}(t\lambda)^{3/2}-y(t\lambda)^{1/2}}&0\\
0&0&e^{-\frac{2}{3}(t\lambda)^{3/2}+y(t\lambda)^{1/2}}
\end{pmatrix}
\end{split}
\end{equation}

2. Across the rays $\arg\lambda=\frac{2\pi}{3}(j-1)$, $j=1,2,3$, oriented
towards infinity, $\Psi_n^{RH}(\lambda)$ has the jumps
\begin{equation}\label{Psi_RH_jumps}
\Psi_n^{RH+}(\lambda)=\Psi_n^{RH-}(\lambda)S_j,\quad
\arg\lambda=\frac{2\pi}{3}(j-1),
\end{equation}
where plus and minus indicate the limiting values of
$\Psi_n^{RH}(\lambda)$ on the jump contour from the left and from the right,
respectively, and
$$
S_1=\begin{pmatrix}
1&2(g_1^{(2)}-g_0^{(2)})(g_2^{(1)}-g_1^{(1)})&
2(g_1^{(2)}-g_2^{(2)})(g_1^{(1)}-g_2^{(1)})\\
0&1&0\\
0&-1&1
\end{pmatrix},
$$
$$
S_2=\begin{pmatrix}
1&2(g_0^{(2)}-g_1^{(2)})(g_2^{(1)}-g_0^{(1)})&
2(g_0^{(2)}-g_2^{(2)})(g_2^{(1)}-g_0^{(1)})\\
0&1&1\\
0&0&1
\end{pmatrix},
$$
$$
S_3=\begin{pmatrix}
1&2(g_1^{(2)}-g_2^{(2)})(g_1^{(1)}-g_0^{(1)})&
2(g_2^{(2)}-g_0^{(2)})(g_1^{(1)}-g_0^{(1)})\\
0&1&0\\
0&-1&1
\end{pmatrix},
$$
and across the ray $\arg\lambda=-\frac{\pi}{3}$ oriented towards infinity,
the following jump condition holds,
\begin{equation}\label{Psi_branch_jump}
\Psi_n^{RH+}(\lambda)=\Psi_n^{RH-}(\lambda)\Sigma,\quad
\arg\lambda=-\frac{\pi}{3},\quad
\Sigma=\begin{pmatrix}
1&0&0\\
0&0&-1\\
0&1&0
\end{pmatrix}.
\end{equation}
\end{RHP}

The wave functions $\psi_n(\lambda)$, $\psi_{n-1}(\lambda)$ and
$\psi_{n-2}(\lambda)$ are just entries of the first column of
$\Psi_n^{RH}(\lambda)$.

\begin{RHP}\label{RHP_phi}
Find a piece-wise holomorphic $3\times3$ matrix function 
$\Phi_m^{RH}(\xi)$ with the following properties:

1. As $\xi\to\infty$,
\begin{equation}\label{Phi_m_as_RH}
\begin{split}
\Phi_m^{RH}(\xi)\to&
\begin{pmatrix}
\frac{\xi^m}{h_m}
&-\frac{th_m}{2\pi^{3/2}}(-1)^m(t\xi)^{-\frac{m+1}{2}-\frac{1}{4}}
&\frac{ith_m}{2\pi^{3/2}}(t\xi)^{-\frac{m+1}{2}-\frac{1}{4}}
\\
\frac{\xi^{m-1}}{h_{m-1}}
&-\frac{th_{m-1}}{2\pi^{3/2}}
(-1)^{m-1}(t\xi)^{-\frac{m}{2}-\frac{1}{4}}
&\frac{ith_{m-1}}{2\pi^{3/2}}(t\xi)^{-\frac{m}{2}-\frac{1}{4}}
\\
\frac{\xi^{m-2}}{h_{m-2}}
&-\frac{th_{m-2}}{2\pi^{3/2}}
(-1)^{m-2}(t\xi)^{-\frac{m-1}{2}-\frac{1}{4}}
&\frac{ith_{m-2}}{2\pi^{3/2}}
(t\xi)^{-\frac{m-1}{2}-\frac{1}{4}}
\end{pmatrix}\times
\\
&\times
\begin{pmatrix}
e^{-\frac{1}{3}\xi^3-y\xi}&0&0\\
0&e^{\frac{2}{3}(t\xi)^{3/2}-x(t\xi)^{1/2}}&0\\
0&0&e^{-\frac{2}{3}(t\xi)^{3/2}+x(t\xi)^{1/2}}
\end{pmatrix}
\end{split}
\end{equation}

2. Across the rays $\arg\xi=\frac{2\pi}{3}(j-1)$, $j=1,2,3$, oriented
towards infinity, $\Phi_m^{RH}(\xi)$ has the jumps
\begin{equation}\label{Phi_RH_jumps}
\Phi_m^{RH+}(\xi)=\Phi_m^{RH-}(\xi)T_j,\quad
\arg\xi=\frac{2\pi}{3}(j-1),
\end{equation}
where plus and minus indicate the limiting values of
$\Phi_m^{RH}(\xi)$ on the jump contour from the left and from the right,
respectively, and
$$
T_1=\begin{pmatrix}
1&2(g_1^{(1)}-g_0^{(1)})(g_2^{(2)}-g_1^{(2)})&
2(g_1^{(1)}-g_2^{(1)})(g_1^{(2)}-g_2^{(2)})\\
0&1&0\\
0&-1&1
\end{pmatrix},
$$
$$
T_2=\begin{pmatrix}
1&2(g_0^{(1)}-g_1^{(1)})(g_2^{(2)}-g_0^{(2)})&
2(g_0^{(1)}-g_2^{(1)})(g_2^{(2)}-g_0^{(2)})\\
0&1&1\\
0&0&1
\end{pmatrix},
$$
$$
T_3=\begin{pmatrix}
1&2(g_1^{(1)}-g_2^{(1)})(g_1^{(2)}-g_0^{(2)})&
2(g_2^{(1)}-g_0^{(1)})(g_1^{(2)}-g_0^{(2)})\\
0&1&0\\
0&-1&1
\end{pmatrix},
$$
and across the ray $\arg\lambda=-\frac{\pi}{3}$ oriented towards infinity,
the following jump condition holds,
\begin{equation}\label{Phi_branch_jump}
\Phi_m^{RH+}(\xi)=\Phi_m^{RH-}(\xi)\Sigma,\quad
\arg\xi=-\frac{\pi}{3},\quad
\Sigma=\begin{pmatrix}
1&0&0\\
0&0&-1\\
0&1&0
\end{pmatrix}.
\end{equation}
\end{RHP}

The wave functions $\phi_n(\lambda)$, $\phi_{n-1}(\lambda)$ and
$\phi_{n-2}(\lambda)$ are just entries of the first column of
$\Phi_n^{RH}(\xi)$.

\bibliographystyle{plain}
\ifx\undefined\bysame
\newcommand{\bysame}{\leavevmode\hbox to3em{\hrulefill}\,}
\fi

\end{document}